\def\ra{\rangle}
\def\la{\langle}
\def\be{\begin{equation}}
\def\ee{\end{equation}}
\def\ba{\begin{array}}
\def\ea{\end{array}}

\def\Cb{{\Bbb C}}

\documentclass[prl,showpacs,twocolumn,amsmath]{revtex4}
\input amssym.def
\begin{document}
\title{Gisin's Theorem for Arbitrary Dimensional Multipartite States}
\author{Ming Li$^{1}$}
\author{Shao-Ming Fei$^{2,3}$}
\affiliation{$^1$Department of Mathematics, China University
of Petroleum, 257061 Dongying, China\\
$^2$Department of Mathematics, Capital Normal University, Beijing
100037, China\\
$^3$Max-Planck-Institute for Mathematics in the Sciences, 04103
Leipzig, Germany}

\begin{abstract}
We present a set of Bell inequalities which are sufficient and
necessary for separability of general pure multipartite quantum states
in arbitrary dimensions. The relations between Bell inequalities
and distillability are also studied. We show that
any quantum states that violate one of these Bell inequalities are distillable.
\end{abstract}

\maketitle

\section{Introduction}

One of the most remarkable aspects of quantum theory is the
incompatibility of quantum non-locality with local-realistic
theories. The Bell inequalities \cite{bell} impose constraints on
the correlations between measurement outcomes on two separated
systems, giving rise to the limits for what can be described within
the framework of any local hidden variable theory. They are of great
importance for understanding the conceptual foundations of quantum
theory as well as for investigating quantum entanglement, as Bell
inequalities can be violated by quantum entangled states. One of the
most important Bell inequalities is the Clauser-Horne-Shimony-Holt
(CHSH) inequality \cite{clauser} for two-qubit systems. It is then
generalized to the $N$-qubit case, known as the Mermin-Ardehali-Belinskii-Klyshko
(MABK) inequality \cite{mermin}. A set of multipartite Bell inequalities has been
elegantly derived in terms of two dichotomic observables per site
\cite{wwzb}, which includes the MABK inequality as a special case
\cite{10} and can detect some entangled states that the MABK
inequality fails to detect. In \cite{chen} another family of Bell
inequalities for $N$-qubit systems has been introduced, which is
maximally violated by all the Greenberger-Horne-Zeilinger states.

In fact, Gisin presented a theorem in 1991. It says that any pure entangled
two-qubit states violate the CHSH
inequality \cite{gisin}. Namely the CHSH inequality is both sufficient
and necessary for separability of two-qubit states.
Soon after, Gisin and Peres provided an elegant
proof of this theorem for the case of pure two-qudit systems \cite{gisinperes}.
In \cite{jingling} Chen et al. showed that all pure entangled three-qubit states
violate a Bell inequality. Nevertheless generally it still remains open whether
the Gisin's theorem can be generalized to the $N$-qudit case or not.

The Bell inequalities are also useful in verifying the security of quantum key
distribution protocols \cite{security}. There is a
simple relation between nonlocality and distillability: if any
two-qubit \cite{ho} or three-qubit \cite{loo} pure or mixed state
violates a specific Bell inequality, then the state must be
distillable. In \cite{16} D\"{u}r has shown that for the case $N\geq 8$,
there exist $N$-qubit bound entangled (not distillable) states which
violate some Bell inequalities. However, Ac\'{i}n has demonstrated
that for all states violating the inequality, there exists at least one
kind of bipartite decomposition of the system such that pure
entangled state can be distilled \cite{18,19}.
But generally it is still an open problem if violation of a Bell
inequality already implies distillability.

In this paper, we present a set of Bell inequalities which can be
shown to be both sufficient and necessary for separability
of general pure multipartite quantum states in arbitrary dimensions, thus
proving the Gisin's theorem generally. We also show that
pure entangled states can be distilled from quantum mixed states that
violate one of these Bell inequalities.

\section{Bell Inequalities for Bipartite Quantum Systems}

For two-qubit quantum systems, the Bell operators are defined by
\be
{\mathcal {B}}=A_1\otimes B_1+A_1\otimes B_2+A_2\otimes B_1-A_2\otimes B_2,
\ee
where
$A_i=\vec{a_i}\cdot \vec{\sigma}_A=a_i^x\sigma_A^x+a_i^y\sigma_A^y+a_i^z\sigma_A^z$,
$B_j=\vec{b_j}\cdot \vec{\sigma}_B=b_j^x\sigma_B^x+b_j^y\sigma_B^y+b_j^z\sigma_B^z$,
$\vec{a_i}=(a_i^x,a_i^y,a_i^z)$ and $\vec{b_j}=(b_j^x,b_j^y,b_j^z)$
are real unit vectors satisfying $|\vec{a_i}|=|\vec{b_j}|=1$, $i,j=1,2$,
$\sigma_{A/B}^{x,y,z}$ are Pauli matrices. The CHSH inequality says that if there exist local hidden
variable models to describe the system, the inequality
\begin{eqnarray}\label{b22}
|\la{\mathcal {B}} \ra|\leq2
\end{eqnarray}
must hold.

In stead of two-qubit ($2\times 2$) system,
we first consider general $N\times M$ bipartite quantum systems in vector space ${\mathcal
{H_{AB}}}={\mathcal {H_A}}\otimes {\mathcal {H_B}}$ with
dimensions $dim\,{\mathcal {H_A}}=M$ and $dim\,{\mathcal {H_B}}=N$
respectively. We aim to find Bell inequalities like (\ref{b22}) such that
any quantum entangled states would violate a Bell inequality.

Let $L_\alpha^A$ and $L_\beta^B$ be the generators of special
unitary groups $SO(M)$ and $SO(N)$ respectively. The $M(M-1)/2$ generators
$L_\alpha^A$ are given by
$\{|j\rangle\langle k|-|k\rangle\langle j|\}$, $1\leq j < k \leq M$,
where $|i\ra$, $i=1,...,M$, are the usual orthonormal basis of ${\mathcal {H_A}}$.
$L_\beta^B$ are similarly defined.
The matrix operators $L_{\alpha}$ (resp. $L_{\beta}$) have $M-2$
(resp. $N-2$) rows and $M-2$ (resp. $N-2$) columns that are
identically zero. We define the operators $A_i^{\alpha}$ (resp.
$B_j^{\beta}$) from $L_{\alpha}$ (resp. $L_{\beta}$) by replacing
the four entries on the positions of the nonzero 2 rows and 2
columns of $L_{\alpha}$ (resp. $L_{\beta}$) with the corresponding
four entries of the matrix $\vec{a_i}\cdot\vec{\sigma}$ (resp.
$\vec{b_j}\cdot\vec{\sigma}$), and keeping the other entries of
$A_i^{\alpha}$ (resp. $B_j^{\beta}$) zero. We define the Bell
operators to be \be\label{bellop} {\mathcal
{B}}_{\alpha\beta}=\tilde{A}_1^{\alpha}\otimes
\tilde{B}_1^{\beta}+\tilde{A}_1^{\alpha}\otimes
\tilde{B}_2^{\beta}+\tilde{A}_2^{\alpha}\otimes
\tilde{B}_1^{\beta}-\tilde{A}_2^{\alpha}\otimes \tilde{B}_2^{\beta},
\ee where
$\tilde{A}_i^{\alpha}=L_{\alpha}A_{i}^{\alpha}L_{\alpha}^{\dag}$,
$\tilde{B}_j^{\beta}=L_{\beta}B_j^{\beta}L_{\beta}^{\dag}$, and $i,
j=1, 2$.

{\bf Theorem 1:} Any bipartite pure quantum state is entangled if
and only if at least one of the following Bell inequalities is
violated,
\be\label{b}
|\la{\mathcal{B}}_{\alpha\beta}\ra|\leq 2,
\ee
where $\alpha=1,2,\cdots, \frac{M(M-1)}{2}$, $\beta=1,2,\cdots,\frac{N(N-1)}{2}$.

{\bf Proof:} Assume that the state $|\psi\ra$ violates one of the
Bell inequalities in (\ref{b}), i.e. there exist $\alpha_0$ and
$\beta_0$ such that $|\la{\mathcal {B}}_{\alpha_0\beta_0}\ra|>2$.
Then equivalently one has that the state
$|\psi\ra_{\alpha_0\beta_0}=\frac{L_{\alpha_0}^A\otimes
L_{\beta_0}^B|\psi\ra}{||L_{\alpha_0}^A\otimes
L_{\beta_0}^B|\psi\ra||}$ violates the CHSH inequality in
(\ref{b22}). As the local operation $L_{\alpha_0}^A\otimes
L_{\beta_0}^B$ does not change the separability of a
state, $|\psi\ra$ must be entangled.

Now assume that $|\psi\ra\in {\mathcal {H_{AB}}}$ is an entangled
state. We prove that at least one of the Bell
inequalities in (\ref{b}) is violated.
Set $\rho=|\psi\ra\la\psi|$.
By projecting $|\psi\ra$ onto $2\times 2$ subsystems \cite{ou},
we get the following pure states:
\begin{eqnarray}\label{qus}
\rho_{\alpha\beta}=\frac{L_{\alpha}^A\otimes
L_{\beta}^B\rho(L_{\alpha}^A)^{\dag}\otimes
(L_{\beta}^B)^{\dag}}{||L_{\alpha}^A\otimes
L_{\beta}^B\rho(L_{\alpha}^A)^{\dag}\otimes (L_{\beta}^B)^{\dag}||},
\end{eqnarray}
where $\alpha=1,2,\cdots, \frac{M(M-1)}{2}; \beta=1,2,\cdots,
\frac{N(N-1)}{2}$, and $||X||=\sqrt{Tr(XX^{\dag})}$.
Here $\rho_{\alpha\beta}$ are pure states with rank one.
As the matrix $L_{\alpha}^A\otimes L_{\beta}^B$ has $MN-4$ rows and
$MN-4$ columns that are identically zero, there are at most $4\times4 = 16$ nonzero elements in the
matrix $\rho_{\alpha\beta}$. The states $\rho_{\alpha\beta}$ are called ``two-qubit" states in this sense.

The concurrence of $|\psi\ra$
is defined by $C(|\psi\ra)=\sqrt{2(1-Tr(\rho_A^2))}$ with
$\rho_A=Tr_B(\rho)$ the reduced density matrix of $\rho$ by tracing over
the subsystem $B$ \cite{con}. A pure quantum state $|\psi\ra$ can be generally expressed as
$|\psi\ra=\sum\limits_{i=1}^{M}\sum\limits_{j=1}^{N}a_{ij}|ij\ra,~
a_{ij}\in {\Bbb{C}}$, in the computational basis $|i\ra$ and $|j\ra$ of
${\mathcal {H_{A}}}$ and ${\mathcal {H_{B}}}$ respectively, $i= 1,
..., M$ and $j = 1, ..., N$. Therefore the concurrence can be expressed as
\begin{eqnarray}
C(|\psi\ra)=\sqrt{\sum\limits_{\alpha=1}^{M}\sum\limits_{\beta=1}^{N}|C(\rho_{\alpha\beta})|^2},
\end{eqnarray}
where $\rho_{\alpha\beta}$ are defined in (\ref{qus}). Since we
have assumed that $|\psi\ra$ is an entangled quantum state,
$C(|\psi\ra)$ must be not zero, i.e. at least one of the
$\rho_{\alpha\beta}$, say $\rho_{\alpha_0\beta_0}$, has non-zero concurrence, $C(\rho_{\alpha_0\beta_0})> 0$.
As we have discussed above,
$\rho_{\alpha_0\beta_0}$ is actually a ``two-qubit" quantum pure state. It
has been shown in \cite{gisin,gisinperes} that an entangled
two-qubit pure state must violate the Bell inequality
(\ref{b22}). Therefore the inequality
$|\la{\mathcal{B}}_{\alpha_0\beta_0}\ra|\leq 2$ is violated. \hfill \rule{1ex}{1ex}

\section{Bell Inequalities for Multipartite Quantum Systems}

We now generalize the results above to multipartite quantum systems.
For convenience we consider that all the subsystems have the same dimensions.
However, as can be seen from the following, our discussions also apply to
multipartite quantum systems with different dimensions.

Let $H$ denote a $d$-dimensional vector space with basis $|i\ra$,
$i=1,2,...,d$. An $L$-partite pure state in
${H}\otimes\cdots\otimes{H}$ is generally of the form,
\begin{eqnarray}\label{purestate}
|\Psi\ra=\sum\limits_{i_{1},i_{2},\cdots
i_{L}=1}^{d}a_{i_{1},i_{2},\cdots i_{L}}|i_{1},i_{2},\cdots
i_{N}\ra,~ a_{i_{1},i_{2},\cdots i_{L}}\in \Cb.
\end{eqnarray}

Let $\alpha$ and $\alpha^{'}$ (resp.$\beta$ and $\beta^{'}$) be
subsets of the subindices of $a$, associated to the same sub-vector
spaces but with different summing indices. $\alpha$ (or
$\alpha^{'}$) and $\beta$ (or $\beta^{'}$) span the whole space of
the given sub-indix of $a$. A
possible combinations of the indices of $\alpha$ and $\beta$
can be equivalently understood as a kind of bipartite decomposition of
the $L$ subsystems, say part A and part B, containing $m$ and $n=L-m$
subsystems respectively.

For a given bipartite decomposition, we can use the analysis similar
to the bipartite case. Let $L_{\alpha}^A$ and
$L_{\beta}^B$ be the generators of special unitary groups $SO(d^m)$
and $SO(d^n)$.  By projecting $|\Psi\ra$ onto $2\times 2$ subsystems
we have the ``two-qubit" pure states:
\begin{eqnarray}\label{quss}
\rho_{\alpha\beta}^p=\frac{L_{\alpha}^A\otimes
L_{\beta}^B\rho(L_{\alpha}^A)^{\dag}\otimes
(L_{\beta}^B)^{\dag}}{||L_{\alpha}^A\otimes
L_{\beta}^B\rho(L_{\alpha}^A)^{\dag}\otimes (L_{\beta}^B)^{\dag}||},
\end{eqnarray}
where $\alpha=1,2,\cdots, \frac{d^m(d^m-1)}{2}; \beta=1,2,\cdots,
\frac{d^n(d^n-1)}{2}$, $p$ labels the bipartite decompositions of
the $L$ subsystems.

For every pure state $\rho_{\alpha\beta}^p$ we define the corresponding
Bell operators
\be\label{q} {\mathcal
{B}}_{\alpha\beta}^p=\tilde{A}_1^{\alpha}\otimes
\tilde{B}_1^{\beta}+\tilde{A}_1^{\alpha}\otimes
\tilde{B}_2^{\beta}+\tilde{A}_2^{\alpha}\otimes
\tilde{B}_1^{\beta}-\tilde{A}_2^{\alpha}\otimes
\tilde{B}_2^{\beta},
\ee
where
$\tilde{A}_i^{\alpha}=L_{\alpha}^AA_{i}^{\alpha}(L_{\alpha}^A)^{\dag}$
and
$\tilde{B}_j^{\beta}=L_{\beta}^BB_j^{\beta}(L_{\beta}^B)^{\dag}$
are the Hermitian operators similarly defined as in (\ref{bellop}).

{\bf Theorem 2:} Any multipartite pure quantum state is entangled
if and only if at least one of the following inequalities
is violated,
\be
\label{bm} |\la{\mathcal {B}}_{\alpha\beta}^p\ra|\leq 2.
\ee

{\bf Proof:} Obviously, multipartite quantum states that violate any
one of the Bell inequalities in (\ref{bm}) must be entangled.

We now prove that, for any entangled multipartite pure quantum
state, at least one of the inequalities in (\ref{bm}) is
violated. The concurrence of $|\Psi\ra$ is given by \cite{anote}
\be\label{def}
\ba{l}
C_{d}^{L}(|\Psi\ra)=\\[3mm]
\sqrt{K\sum\limits_{p}
\sum\limits_{\{\alpha,\alpha^{'},\beta,\beta^{'}\}}^{d}
|a_{\alpha\beta}a_{\alpha^{'}\beta^{'}}-a_{\alpha\beta^{'}}a_{\alpha^{'}\beta}|^{2}},
\ea
\ee
where $K={d}/{2m(d-1)}$, $m=2^{L-1}-1$, $\sum\limits_{p}$ stands for the summation over
all possible combinations of the indices of $\alpha$ and $\beta$.
(\ref{def}) can be rewritten as
\begin{eqnarray}\label{deff}
C_d^L(|\Psi\ra)=\sqrt{K\sum_{p}\sum_{\alpha\beta}(C(\rho_{\alpha\beta}^{p}))^{2}},
\end{eqnarray}
where $\rho^p_{\alpha\beta}$ are defined in (\ref{quss}).
As $|\Psi\ra$ is an entangled state,
$C(|\Psi\ra)$ must be not zero, i.e. at least one of
$\rho_{\alpha\beta}^p$, say $\rho_{\alpha_0\beta_0}^{p_0}$, has non-zero concurrence.
As we have discussed above,
$\rho_{\alpha_0\beta_0}^{p_0}$ is actually a two-qubit quantum pure
state. An entangled two-qubit quantum pure state must violate the Bell
inequality (\ref{b22}). \hfill \rule{1ex}{1ex}

As an example, we consider three-qubit systems.
In \cite{acin}, Acin etc. have verified that any pure three-qubit
state $|\Psi\rangle$ can be uniquely written as
\be
\ba{rcl} |\Psi\rangle &=& \lambda_0
|000\rangle+\lambda_1 e^{i\psi} |100\rangle\\[1mm]
&&+\lambda_2 |101\rangle
+\lambda_3 |110\rangle+\lambda_4 |111\rangle,
\label{gschmidt}
\ea
\ee
where $\lambda_i \geq 0$, $ 0 \leq \psi \leq
\pi$, $\sum_i \lambda_i^2=1$. From straightforward
computation one has
\begin{eqnarray*}\label{3qc}
 C^2(|\Psi\ra)&=&2(\lambda_0\lambda_2)^2+2(\lambda_0\lambda_4)^2
 +|2e^{i\psi}\lambda_1\lambda_4-2\lambda_2\lambda_3|^2\\
 &+&2(\lambda_0\lambda_3)^2+2(\lambda_0\lambda_4)^2
 +|2e^{i\psi}\lambda_1\lambda_4-2\lambda_2\lambda_3|^2\\
 &+&2(\lambda_0\lambda_2)^2+2(\lambda_0\lambda_3)^2+2(\lambda_0\lambda_4)^2.
\end{eqnarray*}
We give a detailed analysis on that an entangled pure three-qubit state, i.e.
at least one of the terms in the right hand side of (\ref{3qc})
is non-zero, must violate one of the inequalities in (\ref{bm}).

{\bf{Case 1:}} If $\lambda_0\lambda_2\neq 0$, the corresponding operator\\
$L_2^A\otimes L_1^B=\left(%
    \begin{array}{cccc}
      0 & 0 & 1 & 0\\
      0 & 0 & 0 & 0\\
      -1 & 0 & 0 & 0\\
      0 & 0 & 0 & 0\\
    \end{array}%
    \right) \otimes \left(%
    \begin{array}{cc}
      0 & 1 \\
      -1 & 0 \\
    \end{array}%
    \right)$ and
$$\ba{l}\rho_{21}^{12|3}=\\[3mm]
\left(%
    \begin{array}{cccccccc}
      \lambda_2^2 & -e^{-i\psi}\lambda_1\lambda_2 & 0 & 0 & 0 & \lambda_0\lambda_2 & 0 & 0\\
      -e^{i\psi}\lambda_1\lambda_2 & \lambda_1^2 & 0 & 0 & 0 & -e^{i\psi}\lambda_0\lambda_1 & 0 & 0\\
      0 & 0 & 0 & 0 & 0 & 0 & 0 & 0\\
      0 & 0 & 0 & 0 & 0 & 0 & 0 & 0\\
      0 & 0 & 0 & 0 & 0 & 0 & 0 & 0\\
      \lambda_0\lambda_2 & -e^{-i\psi}\lambda_0\lambda_1 & 0 & 0 & 0 & \lambda_0^2 & 0 & 0\\
      0 & 0 & 0 & 0 & 0 & 0 & 0 & 0\\
      0 & 0 & 0 & 0 & 0 & 0 & 0 & 0\\
    \end{array}%
    \right).
    \ea
    $$
Choose the Bell operator in (\ref{q}) to be the one with respect to the bipartite decomposition
of the first two qubits and the last one,
    \be {\mathcal
{B}}_{21}^{12|3}=\tilde{A}_1^{2}\otimes
\tilde{B}_1^{1}+\tilde{A}_1^{2}\otimes
\tilde{B}_2^{1}+\tilde{A}_2^{2}\otimes
\tilde{B}_1^{1}-\tilde{A}_2^{2}\otimes \tilde{B}_2^{1},
\ee
where
$\tilde{A}_k^{2}=L_{2}^AA_{k}^{2}(L_{2}^A)^{\dag}$, $\tilde{B}_l^{1}=L_{1}^BB_l^{1}(L_{1}^B)^{\dag}$,
and $A_k^2=\left(%
    \begin{array}{cccc}
      -a_k^3 & 0 & a_k^1+a_k^2i & 0\\
      0 & 0 & 0 & 0\\
      a_k^1-a_k^2i & 0 & a_k^3 & 0\\
      0 & 0 & 0 & 0\\
    \end{array}%
    \right)$, $B_l^1=\left(%
    \begin{array}{cccc}
      -b_l^3 & b_l^1+b_l^2i\\
      b_l^1-b_l^2i & b_l^3\\
      \end{array}%
    \right)$, $k,l=1,2$,
we have the maximal violation of the inequality (\ref{bm}),
$2\sqrt{1+\frac{4\lambda_0^2\lambda_2^2}{(\lambda_0^2+\lambda_1^2+\lambda_2^2)^2}}>2$.

{\bf{Case 2:}} If
$|e^{i\psi}\lambda_1\lambda_4-\lambda_2\lambda_3|\neq 0$, the
corresponding operator
$L_6^A\otimes L_1^B=\left(%
    \begin{array}{cccc}
      0 & 0 & 0 & 0\\
      0 & 0 & 0 & 0\\
      0 & 0 & 0 & 1\\
      0 & 0 & -1 & 0\\
    \end{array}%
    \right) \otimes \left(%
    \begin{array}{cc}
      0 & 1 \\
      -1 & 0 \\
    \end{array}%
    \right)$. The matrix $\rho_{61}^{12|3}$ has only nonzero entries at the right down corner with the form,
    $$\left(
    \begin{array}{cccc}
      \lambda_4^2 & -\lambda_3\lambda_4 & -\lambda_2\lambda_4 & e^{-i\psi}\lambda_1\lambda_4 \\
      -\lambda_3\lambda_4 & \lambda_3^2 & \lambda_2\lambda_3 & -e^{-i\psi}\lambda_1\lambda_3\\
      -\lambda_2\lambda_4 & \lambda_2\lambda_3 & \lambda_2^2 & -e^{-i\psi}\lambda_1\lambda_2\\
      e^{i\psi}\lambda_1\lambda_4 & -e^{i\psi}\lambda_1\lambda_3
      & -e^{i\psi}\lambda_1\lambda_2 & \lambda_1^2\\
    \end{array}%
    \right).$$
The Bell operator in (\ref{q}) has the form,
    \be {\mathcal
{B}}_{61}^{12|3}=\tilde{A}_1^{6}\otimes
\tilde{B}_1^{1}+\tilde{A}_1^{6}\otimes
\tilde{B}_2^{1}+\tilde{A}_2^{6}\otimes
\tilde{B}_1^{1}-\tilde{A}_2^{6}\otimes \tilde{B}_2^{1},\ee
where
$\tilde{A}_k^{6}=L_{6}^AA_{k}^{6}(L_{6}^A)^{\dag}$, $\tilde{B}_l^{1}
=L_{1}^BB_l^{1}(L_{1}^B)^{\dag}$, and $A_k^6=\left(%
    \begin{array}{cccc}
      0 & 0 & 0 & 0\\
      0 & 0 & 0 & 0\\
      0 & 0 & -a_k^3 & a_k^1+a_k^2i\\
      0 & 0 & a_k^1-a_k^2i & a_k^3\\
    \end{array}%
    \right)$, $B_l^1=\left(%
    \begin{array}{cccc}
      -b_l^3 & b_l^1+b_l^2i\\
      b_l^1-b_l^2i & b_l^3\\
      \end{array}%
    \right)$, $k,l=1,2.$
The corresponding maximal violation is given by
$2\sqrt{1+\frac{4|e^{i\psi}\lambda_1\lambda_4-\lambda_2\lambda_3|^2}
{(\lambda_1^2+\lambda_2^2+\lambda_3^2+\lambda_4^2)^2}}$,
which is obviously strictly larger than 2.
Other cases can be discussed similarly.

\section{Bell Inequalities and Distillation}

A bipartite state $\rho$ is called distillable, iff maximally entangled
bipartite pure states, e.g.
$|\Phi^+\ra=\frac{1}{\sqrt{2}}(|00\ra+|11\ra)$, can be created
from a number of identical copies of the state $\rho$ by means of
local operations and classical communication. We
call a multipartite state distillable,  if and only if there exists at least
one bipartite decomposition of the system such that pure entangled states
can be distilled. It has been shown that all quantum entangled
pure states are distillable. However it is a challenge to give an operational
criterion of distillability for general mixed states. In \cite{ou} a
sufficient condition of distillability has been presented. Our inequalities
(\ref{bm}) are both sufficient and necessary for separability of pure states,
but generally not for separability of mixed ones.
However surprisingly (\ref{bm}) can be served as
criterion for distillability.

{\bf Theorem 3:} Any bipartite quantum state $\rho$ that violates any one of the
Bell inequalities in (\ref{b}), i.e. Tr$\{{\mathcal
{B}}_{\alpha\beta}\rho\}>2$, is always distillable.
And if a multipartite quantum state $\rho$ violates one of the
Bell inequalities in (\ref{bm}), i.e. $\rho$ satisfies Tr$\{{\mathcal
{B}}_{\alpha\beta}^p\rho\}>2$, then bipartite maximally
entangled pure states can be distilled from the copies of $\rho$.

{\bf Proof:} It was shown in \cite{horo} that a density matrix
$\rho$ is distillable iff there are some projectors $P$, $Q$ that
map high dimensional spaces to two-dimensional ones such that the
state $P\otimes Q \rho^{\otimes s} P\otimes Q$ is entangled for some
$s$ copies. Thus if any one of the Bell inequalities in (\ref{b}) is
violated, there exists a submatrix $\rho_{\alpha\beta}$, like
(\ref{qus}), that has nonzero concurrence.
For generally given operator $L_\alpha=|i\rangle\langle j|-|j\rangle\langle i|$,
$L_\beta=|k\rangle\langle l|-|l\rangle\langle k|$,
the operators $P$, $Q$ can be explicitly given by
$P=A L_\alpha$, $Q=B L_\beta$, where
$A=|0_A\ra\la i|+|1_A\ra\la j|$, $B=|0_B\ra\la
k|+|1_B\ra\la l|$, $|0_{A/B}\ra$ and $|1_{A/B}\ra$ are
the orthonormal bases of a two dimensional vector space.
$P\otimes Q$ maps state $\rho$ to a two-qubit one that
has the same nonzero concurrence as $\rho_{\alpha\beta}$.
Since any entangled two-qubit state is distillable, $\rho$ is
distillable. The multipartite case can be discussed similarly.
\hfill \rule{1ex}{1ex}

{\it Remark} It has been shown that PPT (positive partial
transposition) entangled quantum states are not distillable \cite{ppt}.
Therefore PPT quantum states should
never violate the Bell inequalities in (\ref{b}) or (\ref{bm}).
This fact can be seen from the following.
A density matrix $\rho$ is called PPT if the partial transposition
of $\rho$ with respect to any subsystem(s) is still positive. Let
$\rho^{T_{B}}$ denote the partial transposition with respect to the
subsystem $B$. Assume that there is a PPT state $\rho$ violating
one of the Bell inequalities in (\ref{bm}), say Tr$\{{\mathcal
{B}}_{\alpha_0\beta_0}^{p_0}\rho\}>2$. This can be equivalently
understood as that there exists two-qubit state
$\rho_{\alpha_{0}\beta_{0}}^{p_0}$ in the form of (\ref{quss}) such
that
Tr$\{B_{\alpha_0\beta_0}^{p_0}\rho_{\alpha_{0}\beta_{0}}^{p_0}\}>2$,
where $B_{\alpha_0\beta_0}^{p_0}=A_1^{\alpha_0}\otimes
B_1^{\beta_0}+A_1^{\alpha_0}\otimes B_2^{\beta_0}+A_2^{\alpha_0}\otimes
B_1^{\beta_0}-A_2^{\alpha_0}\otimes B_2^{\beta_0}$.
One the other hand, by using the PPT property of $\rho$, we have:
\begin{eqnarray}\label{19}
\rho_{\alpha_{0}\beta_{0}}^{T_{B}}=L_{\alpha_{0}}^{A}\otimes
(L_{\beta_{0}}^{B})^{*}\rho^{T_{B}}
(L_{\alpha_{0}}^{A})^{\dag}\otimes (L_{\beta_{0}}^{B})^{T}\geq 0.
\end{eqnarray}
As both $L_{\alpha_{0}}^{A}$ and $ L_{\beta_{0}}^{B}$ are
projectors to two-dimensional subspaces,
$\rho_{\alpha_{0}\beta_{0}}^{p_0}$ can be considered as a $2\times 2$
state. While a $2\times 2$ PPT state
$\rho_{\alpha_{0}\beta_{0}}$ must be separable \cite{ho96}, it
contradicts with Tr$\{B_{\alpha_0\beta_0}^{p_0}\rho_{\alpha_{0}\beta_{0}}^{p_0}\}>2$.

\section{\bf Conclusions and Remarks}

In conclusion, we have derived a series of new Bell inequalities for
both bipartite and multipartite quantum states by projecting the
whole quantum systems to ``two-qubit" subsystems. We show that
quantum states that violating any one of these Bell inequalities
are entangled. On the other hand, we have proved that any entangled
pure quantum states must violate at least one of these Bell
inequalities. Thus the Gisin theorem for general multipartite quantum systems has been
proved. We have also shown that quantum states that violate the Bell
inequalities must be distillable, which helps on
measurable determination of quantum entanglement experimentally.

\noindent{\bf Acknowledgments}\,
This work is supported by the NSFC 10875081,
KZ200810028013 and PHR201007107.

\end{document}